\documentclass[11pt,twoside]{article}
\usepackage{asp2010}

\resetcounters

\begin{document}

\title{Model Atmospheres From Very Low Mass Stars to Brown Dwarfs}
\author{Allard, F.$^1$, Homeier, D.$^2$, and Freytag, B.$^{1,3}$}
\affil{$^1$Centre de Recherche Astrophysique de Lyon, UMR 5574: CNRS, Universit\'e de Lyon, \'Ecole Normale Sup\'erieure de Lyon , 46 All\'ee dÕItalie , F-69364 Lyon Cedex 07, France}
\affil{$^2$Institut f\"ur Astrophysik G\"ottingen, Georg-August-Universit\"at, Friedrich-Hund-Platz 1, 37077 G\"ottingen, Germany}
\affil{$^3$Istituto Nazionale di Astrofisica, Osservatorio Astronomico di Capodimonte, Via Moiariello 16, I-80131~Naples, Italy}

\begin{abstract}
Since the discovery of brown dwarfs in 1994, and the discovery of dust cloud formation in the latest Very Low Mass Stars (VLMs) and Brown Dwarfs (BDs) in 1996, the most important challenge in modeling their atmospheres as become the understanding of cloud formation and advective mixing.  For this purpose, we have developed radiation hydrodynamic 2D model atmosphere simulations to study the formation of forsterite dust in presence of advection, condensation, and sedimentation across the M-L-T VLMs to BDs sequence ($T_\mathrm{eff}=$ 2800\,K to 900\,K, Freytag et al. 2010\nocite{Freytag2010}).  We discovered the formation of gravity waves as a driving mechanism for the formation of clouds in these atmospheres, and derived a rule for the velocity field versus atmospheric depth and $T_\mathrm{eff}$, which is relatively insensitive to gravity.  This rule has been used in the construction of the new model atmosphere grid, {\it\bf BT-Settl}, to determine the micro-turbulence velocity, the diffusion coefficient, and the advective mixing of molecules as a function of depth.  This new model grid of atmospheres and synthetic spectra has been computed for 100,000\,K $> T_\mathrm{eff} >$ 400\,K, 5.5 $>$ logg $>$ -0.5, and [M/H]= +0.5 to -1.5, and the reference solar abundances of \cite{Asplund09}.  We found that the new solar abundances allow an improved (close to perfect) reproduction of the photometric and spectroscopic VLMs properties, and, for the first time, a smooth transition between stellar and substellar regimes --- unlike the transition between the {\it\bf NextGen} models from Hauschildt et al. 1999a,b\nocite{NGa,NGb}, and the {\it\bf AMES-Dusty} models from Allard et al. 2001\nocite{Allard01}).  In the BDs regime, the {\it\bf BT-Settl} models propose an improved explanation for the M-L-T spectral transition.  In this paper, we therefore present the new {\it\bf BT-Settl} model atmosphere grid, which explains the entire transition from the stellar to planetary mass regimes.
\end{abstract}

\section{The Impact of the new Solar Abundances on VLMs spectral properties}
The modeling of the atmospheres of very low mass stars (hereafter VLMs) has evolved with the development of computing capacities from an analytical treatment of the transfer equation using moments of the radiation field \cite[]{AllardPhDT90}, to a line-by-line opacity sampling in spherical symmetry \cite[]{ARAA97,NGa,NGb}, and finally 3D radiation transfer \cite[]{SHB2010}. In parallel to detailed radiative transfer in an assumed static environment, hydrodynamical simulations have been developed to reach a realistic representation of the granulation and the line profiles shifted and shaped by the hydrodynamical flow of the sun and sun-like stars (see for details the review in a special isssue of the Journal of Computational Physics by Freytag et al. 2011) by using a non-grey (multi-group binning of opacities) radiative transfer using a pure blackbody source function (scattering is neglected). 

\begin{figure}[!ht]
\plotone{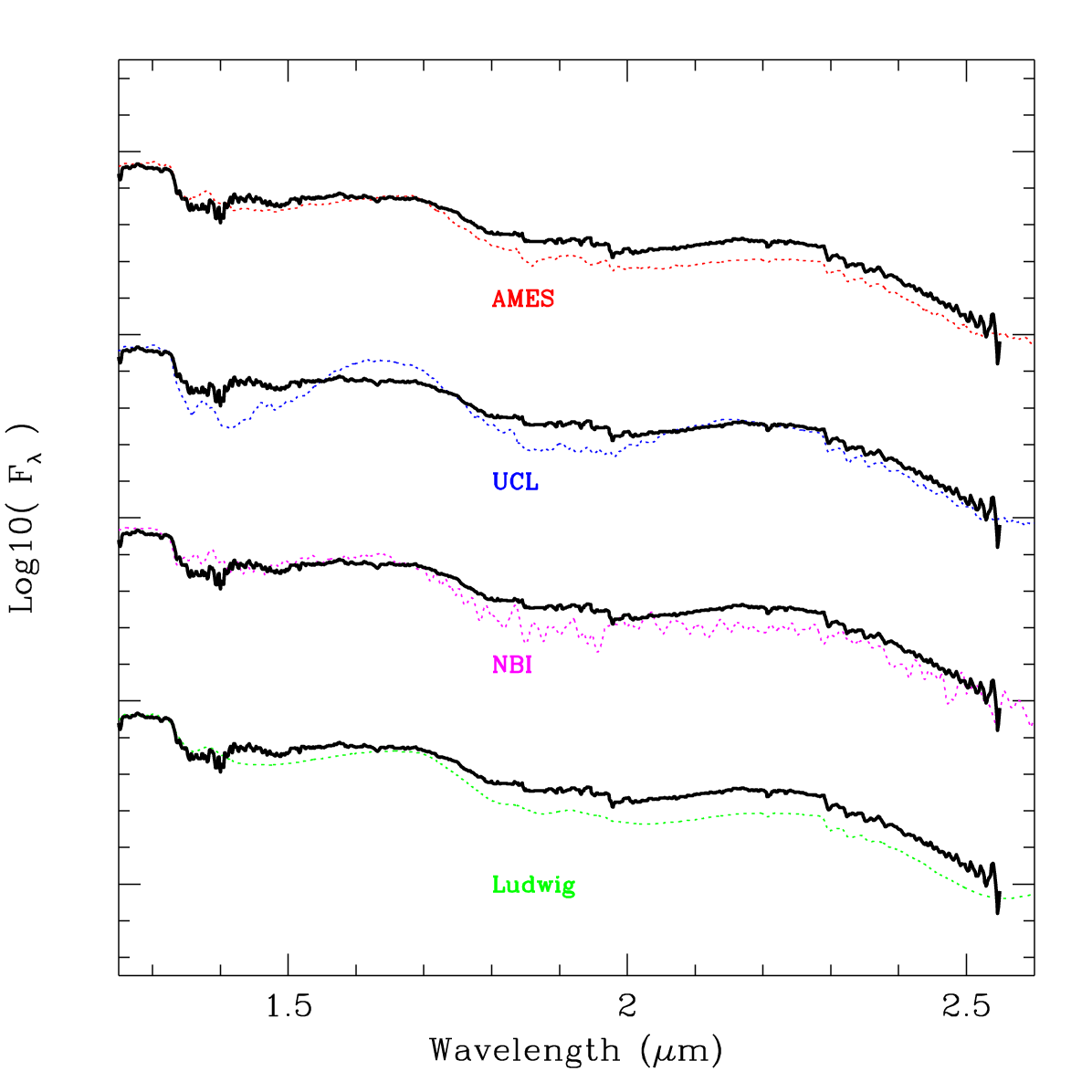}
\caption{Model atmosphere synthetic spectrum versus VB10 IR SED, using different water opacity profiles over the years \cite[]{LudwigH2O,NBIH2O,UCL94H2O,AMESH2O}.}
\label{f:allard_f1_water}
\end{figure}

The model atmospheres and synthetic spectra have also been made possible thanks to the development of realistic opacities calculated often ab initio for the needs of an accurate account of their cooling and heating effects in the internal atmospheric layers where temperatures close to 3000\,K can prevail. For some time, the remaining discrepancies in the model synthetic spectra were believed to be due to incomplete water vapor line lists, both in temperature and in the rotation quantum number {\bf $J$} of the molecular simulations.   In Fig.\ref{f:allard_f1_water}, model atmosphere synthetic spectra are shown using different published water vapor opacity profiles such as hot flames laboratory experiments \cite[]{LudwigH2O}, empirical calculations \cite[]{NBIH2O}, and ab initio calculations \cite[]{UCL94H2O,AMESH2O} based on independently measured interaction potential surface. As can be seen from Fig.\ref{f:allard_f1_water} where the models are compared to the infrared spectrum of an M8 dwarf (VB10),  the water vapor opacity profile which shape this part of the spectrum has strongly change over time with the improvement of computational capacities and a better knowledge of the interaction potential surface. But in general, all opacity profiles converge in predicting an over-opacity (or lack of flux in the model) in the $K$ bandpass.  The UCL opacity profile --- likely because of its incompleteness --- could allow a seemingly correct $J-K$ color while not allowing a detailed spectral comparison of the {\it\bf NextGen} models \cite[]{AllardH2O94}. It became however clear that the more recent versions of the water vapor profile \cite[]{AMESH2O,BT2H2O} all agree in establishing a systematic lack of flux in the $K$ bandpass in the models. This is the reason why the \cite{Allard01} {\it\bf AMES-Cond/Dusty} model grids, based on the \cite{AMESH2O} water vapor opacity profile were never proposed for the study of VLMs, and reserved for the study of the limiting properties of brown dwarfs.  
  
\begin{figure}[!ht]
\plotone{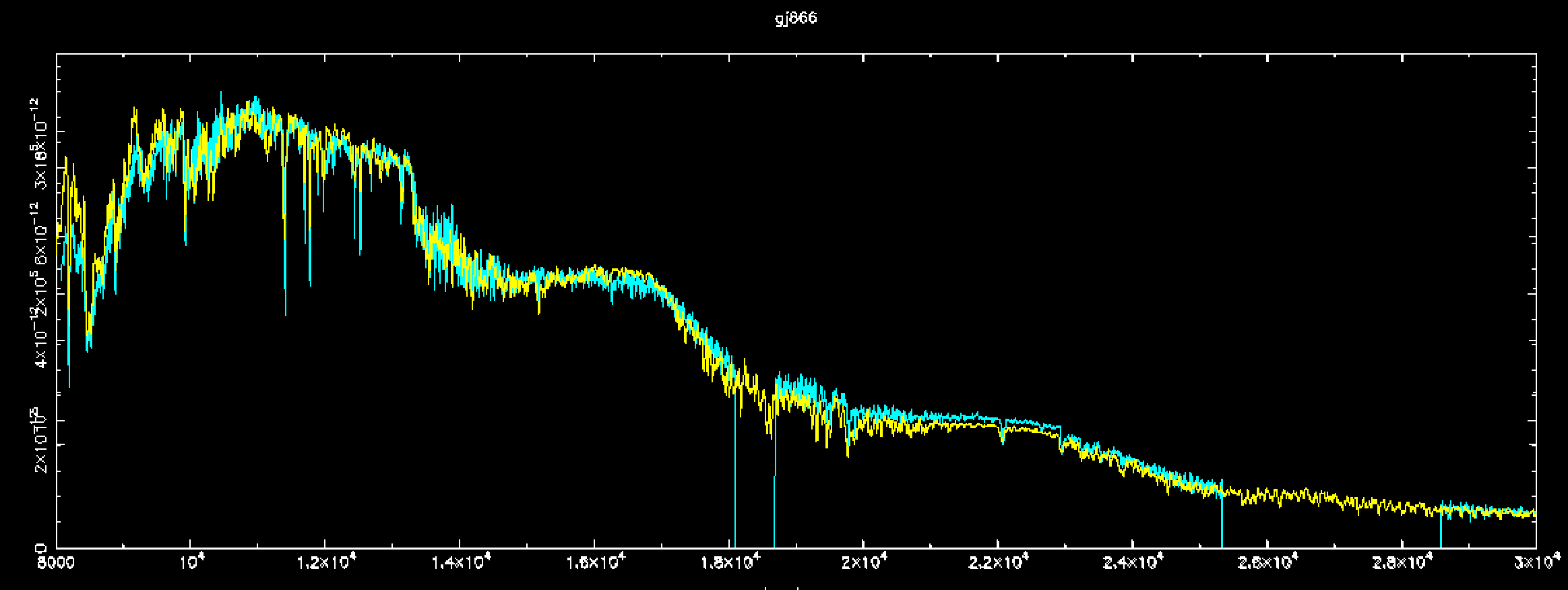}
\plotone{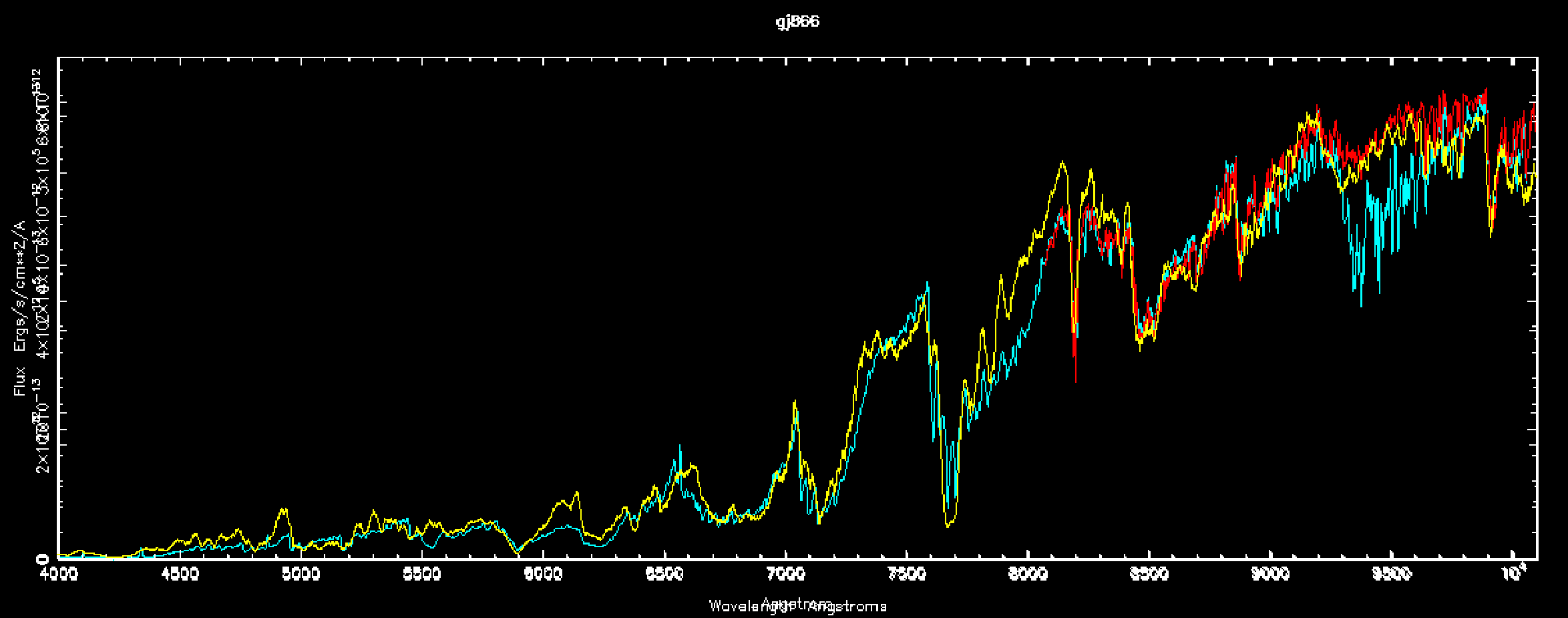}
\caption{A {\it\bf BT-Settl} synthetic spectrum with Teff=2900K, logg=5.0, and solar metallicity by \cite{Asplund09} is shown in yellow compared to the SED of the red dwarf GJ 866 (cyan curve). Both are plotted in absolute flux, i.e no flux adjustment at a specific wavelength is needed. The infrared SED (top panel) is now perfectly reproduced by the model. The agreement is particularly good in the Wing Ford FeH bands near $0.99~{\mu}$m while some discrepancy prevail in the missing CaOH bands around 6000\AA. Telluric absorption is corrected from the observations in red.}
\label{f:allard_f2_GJ866}
\end{figure}

In this regard, the new solar abundances based on radiation hydrodynamical simulations of the solar atmosphere \cite[]{Asplund09,Caffau2010} help as they predict an oxygen abundance of 0.3~dex (a factor of 2) lower than the previously used solar abundances of \cite{GNS93}. The results are shown in Fig.\ref{f:allard_f2_GJ866}, where the new {\it\bf BT-Settl} model atmosphere synthetic spectra for a $T_\mathrm{eff} =$ 2900\,K, logg=5.0 and solar abundances according \cite{Asplund09} are compared to the spectrum of an M6 dwarfs, GJ866 (kindly provided by M. Bessel, Mt-Stromlo Obs.).  For the first time, we find a perfectly fitting spectra distribution across the near-IR to infrared spectral region (the model is the yellow line). The agreement is also excellent in the optical to red part of the spectrum in particularly in the FeH Wing Ford bands near $0.99~{\mu}$m, and in the VO bands thanks to line lists provided by B. Plez (GRAAL, Montpellier, France).  Missing opacities are, however, affecting still the spectral distribution (CaOH bands), and the \cite{AllardTiOH2O2000} TiO line list becomes now too corse compared to progress made with water vapor.

\begin{figure}[!ht]
\plotone{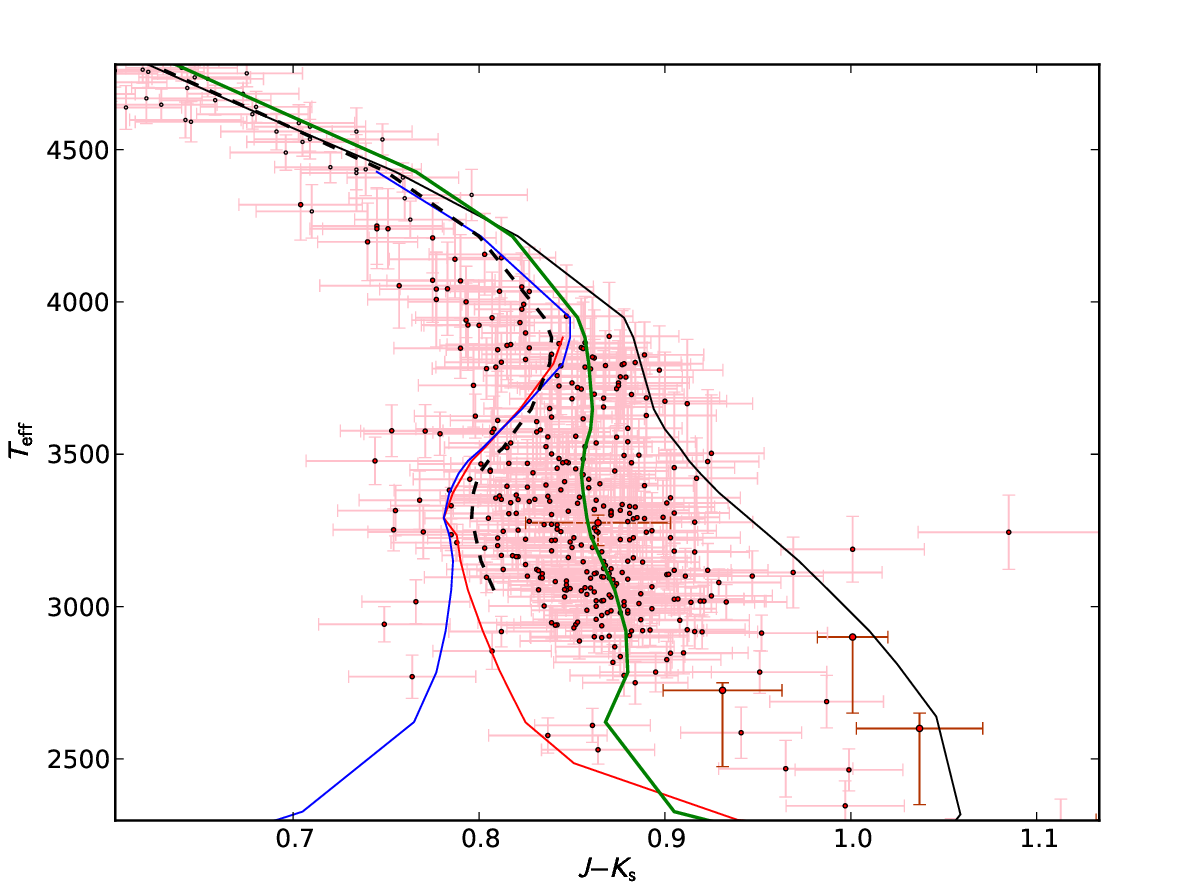}
\caption{Estimated Teff for M dwarfs by \cite{MdwarfsTeff2008} and brown dwarfs by \cite{Golimowski04} are reported as a function of J-K short. Over-plotted are the {\it\bf NextGen} model isochrones for 5 Gyrs \cite[]{BCAH97,BCAH98} using varius generations of model atmospheres, starting with the {\it\bf NextGen} (black line), pursuing with the limiting case {\it\bf AMES-Cond/Dusty}  grids by \cite{Allard01} (blue and red line respectively), and finishing with the {\it\bf BT-Settl} models using the \cite{Asplund09} solar abundances (green line). }
\label{f:allard_f3_Teff-J-Ks}
\end{figure}

One can see from Fig.\ref{f:allard_f3_Teff-J-Ks} that the {\it\bf NextGen} models systematically and increasingly overestimate $T_\mathrm{eff}$ through the lower main sequence, while the {\it\bf AMES-Cond/Dusty} models were underestimating $T_\mathrm{eff}$ compared to the averaged empirical determinations of $T_\mathrm{eff}$ of individual stars \cite[]{MdwarfsTeff2008}. This situation is relieved  when using the newer \cite{Asplund09} abundances, and the {\it\bf BT-Settl} models now agree fairly well with most of the empirical estimations of $T_\mathrm{eff}$. Evolution models are currently being prepared using the {\it\bf BT-Settl} model atmosphere grid.

\section{Dust formation in late type VLMs and Brown Dwarf Atmospheres}

\begin{figure}[!ht]
\plotone{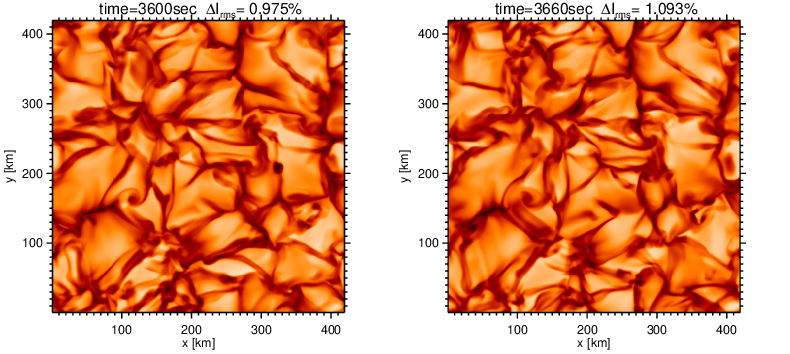}
\plotone{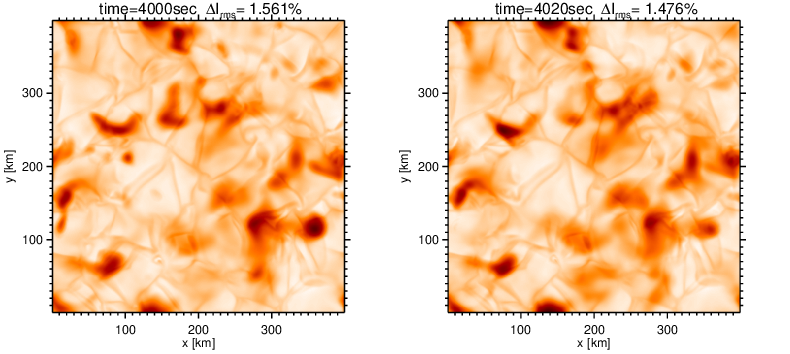}
\plotone{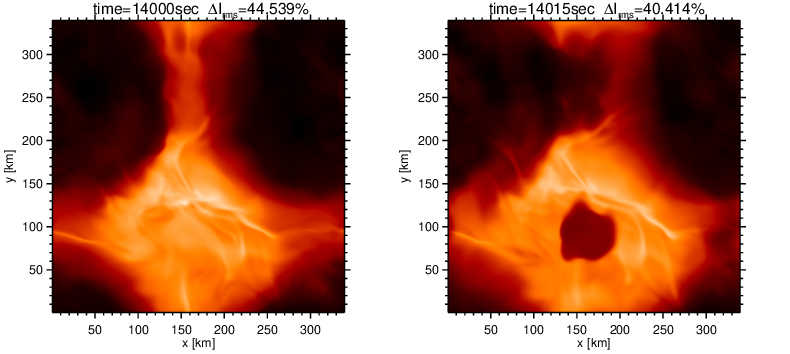}
\caption{3D HRD simulations using {\sl\bf CO5BOLD} \cite[]{Freytag2010} of a $350 \times 350 \times 170$ km$^3$ of atmosphere at the surface of, from top to bottom,  2600K, 2200K and 1500K VLMs and brown dwarfs of logg=5.0, and solar metallicity. For each model two intensity snapshots are shown in time to illustrate the intensity variability. }
\label{f:allard_f4_3DHRD}
\end{figure}

One of the most important challenge in modeling the atmospheres and spectral properties of VLMs and brown dwarfs is the formation of dust clouds and its associated greenhouse effects making the infrared colors of late M and early L dwarfs extremely red compared to colors of low mass stars.    The cloud composition, according to equilibrium chemistry, is going from zirconium oxide (ZrO$_2$), to refractory ceramics (perovskite and corundum; CaTiO$_3$, Al$_2$O$_3$), to silicates (forsterite; Mg$_2$SiO$_4$), to salts (CsCl, RbCl, NaCl), and finally to ices (H$_2$O, NH$_3$, NH$_4$SH) as brown dwarfs cools down with time from M through L, T  spectral types and beyond \cite[]{Allard01,LF06}.    Many cloud models have been constructed to address this problem in brown dwarfs over the past decade \cite[see][for a review on the subject]{Helling08}. However, none treated the mixing properties of the atmosphere, and the resulting diffusion mechanism realistically enough to reproduce the properties of the spectral transition from M through L and T spectral types without changing cloud parameters \cite[for example][]{AM01}.  It is in this context that we have decided to address the issue of mixing and diffusion by 2D Radiation HydroDymanic (hereafter RHD) simulations of VLMs and brown dwarfs atmospheres, using the {\sl\bf Phoenix} opacities in a multi-group binning, and forsterite geometric cross-sections \cite[]{Freytag2010}. We found that gravity waves have a decisive role in clouds formation in brown dwarfs, while around $T_\mathrm{eff} \le$~2200\,K the cloud layers become optically thick enough to initiate cloud convection, which participate in the global mixing. Overshoot is also important in the mixing of the largest dust particles (see paper by D. Homeier in this book).  In Fig.\ref{f:allard_f4_3DHRD}, 3D RHD simulations (a $350 \times 350 \times 170$ km$^3$ box at the surface of the star) are shown for dwarfs with $T_\mathrm{eff} = $ 2600\,K, 2200\,K, and 1500\,K from top to bottom. For each simulation, two snapshots are shown side-by-side to illustrate the intensity variation due to cloud formation and granulation. The 2600\,K case shows no or negligible dust formation, while dust formation progresses to reach optically thick density at around 2200K, before sedimenting out again towards the 1300K regime. The $T_\mathrm{eff} = $1500\,K case illustrate the importance of gravity waves, where the minima of the waves reach condensation levels while the maxima remain in condensed phase. The box of simulations are too small compared to the radius of the star to show adequately the variability of these objects, however. See \cite{Freytag2010} and B. Freytag's poster paper in these proceedings for details.

\begin{figure}[!ht]
\plotone{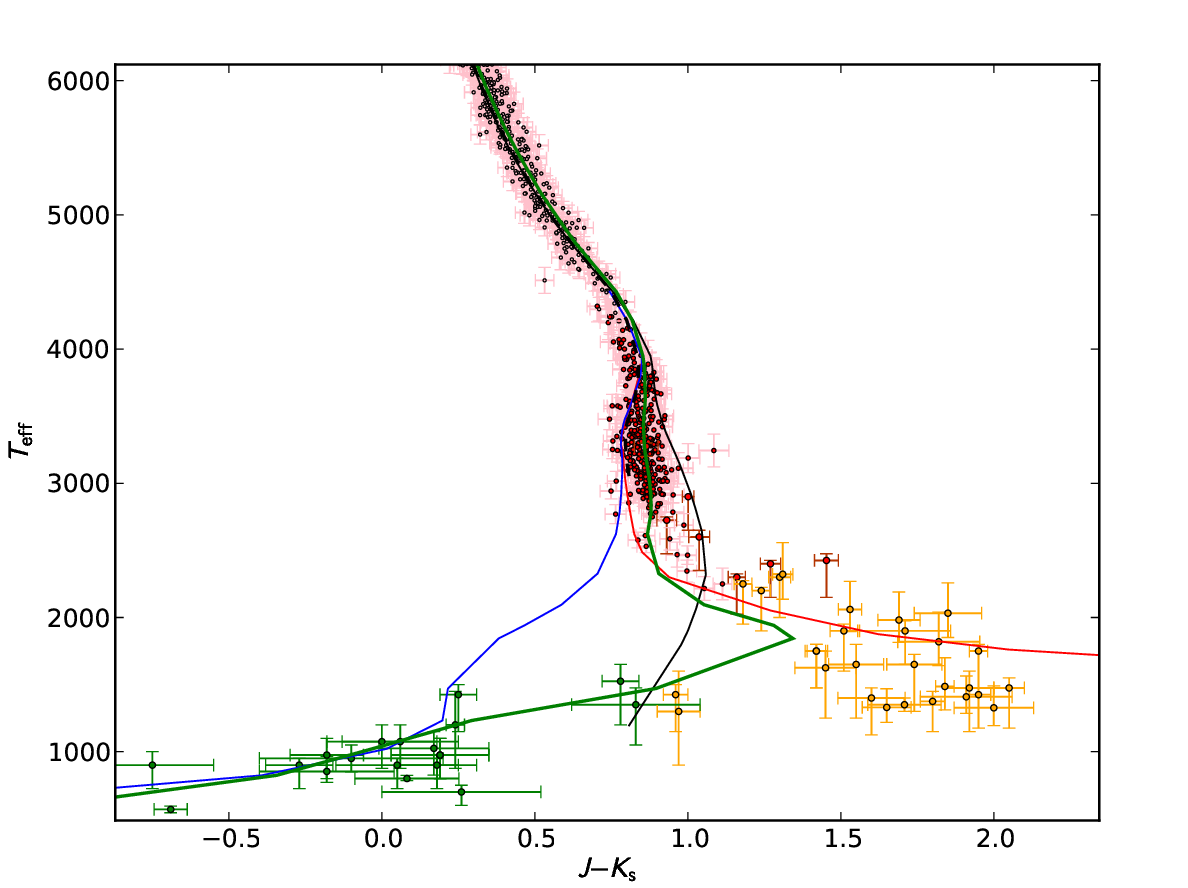}
\caption{Same plot as Fig.\ref{f:allard_f3_Teff-J-Ks} but zooming out and extending into the brown dwarfs region of the diagram. This region below 2500K is dominated by dust formation (essentially forsterite and other silicates).  The limiting case {\it\bf AMES-Cond/Dusty} models atmosphere provide a description of the span in colors of the brown dwarfs in this diagram. }
\label{f:allard_f5_Teff-J-Ks_cool}
\end{figure}

These simulations permitted to account in {\sl\bf Phoenix} for the advective forces bringing fresh condensible material from the hotter lower layers to the cloud forming layers.  Our cloud model is based on the condensation and sedimentation timescales from a study of Earth, Venus, Mars and Jupiter atmospheres by \cite{Rossow78}. However we had to compute the supersaturation pressure from our pre-tabulated equilibrium chemistry in order to obtain the correct amount of dust formation (as opposed to use an approximate value cited in Rossow, 1978\nocite{Rossow78}).  We then solved the cloud model and equilibrium chemistry in turn layer by layer inside out to account for the sequence of grain species formation as a function of cooling of the gas. One can see from Fig.\ref{f:allard_f5_Teff-J-Ks_cool} that the late-type M and early type L dwarfs behave as if dust is formed nearly in equilibrium with the gas phase with extremely red colors in some agreement with the {\it\bf BT-Dusty} models.  At the low Teff regime dominated by T-type dwarfs, the {\it\bf AMES-Cond} models also appear to provide a good limitation for the brown dwarf colors.  However, the chemistry of T dwarfs can be very far from the equilibrium in the models. The {\it\bf BT-Settl} models, which account for a cloud model, dynamical mixing from RHD simulations, a supersaturation computed from pre-tabulated equilibrium chemistry calculations,  and the \cite{Asplund09} solar abundances, manage to reproduce the main sequence down to the L-type brown dwarf regime, before turning to the blue in the late-L and T dwarf regime. The models used an age of 5~Gyrs, and in the case of the {\it\bf BT-Settl} models, a younger age of a few Gyrs would easily reproduce the reddest brown dwarfs.

\section{Summary and Futur Prospects}

We propose, in this paper, a new model atmosphere grid, named {\it\bf BT-Settl}, computed using the atmosphere code {\it{\bf Phoenix}} which has been updated, compared to the \cite{Allard01} {\it\bf AMES-Cond/Dusty} models,  for: i) the \cite{BT2H2O} BT2 water opacity line list, ii) the solar abundances revised by \cite{Asplund09}, and iii) a cloud model accounting for supersaturation and RHD mixing.  It is covering the whole range of VLMs and brown dwarfs and beyond:    1000,000\,K $< T_\mathrm{eff} <$  400\,K;    -0.5 $<$ logg $<$ 5.5; and +0.5 $<$ [M/H] $<$ -4.0, including various values of the alpha element enhancement.  Only the confrontation of the models using spectral synthesis will allow to define the content of oxygen and alpha elements of VLMs and brown dwarfs.  But it is clear that these objects are excellent constraint for the solar oxygen abundance.  The models are available at the Phoenix simulator website "http://phoenix.ens-lyon.fr/simulator/" and are in preparation for publication.  However, the interior and evolution models are expected for the second half of 2011. 

In order to say something about the spectral variability of VLMs and brown dwarfs, 3D RHD simulations of "the star in the box" with rotation will be required. This  is our current project supported by the French "Agence Nationale de la Recherche" for the period 2010-2015.  Rotation is already modeled for a scaled down model of the Sun using {\sl\bf CO5BOLD} \cite[]{Sun_Rotation07} and can be applied to brown dwarf simulations. 

\acknowledgements We would like to thank specifically Mickael Bessel (Mt Stromlo Obs.) for his visit to CRAL and the fruitful discussions, as well as Robert Barber (UCL) for his generous support. We thank the french  "Agence Nationale de la Recherche" (ANR) and "Programme National de Physique Stellaire"  (PNPS) of CNRS (INSU) for their financial support. The computations of dusty M dwarf and brown dwarf models were performed at the {\sl P\^ole Scientifique de Mod\'elisation Num\'erique} (PSMN) at the {\sl \'Ecole Normale Sup\'erieure} (ENS) in Lyon.

\bibliography{allard_f}

\end{document}